\documentclass[prl,aps,twocolumn,showpacs,floatfix]{revtex4-1}
\usepackage{amsfonts}
\usepackage{amssymb}
\usepackage{hyperref}
\usepackage{graphicx}
\usepackage{dcolumn}
\usepackage{bm,amsmath,verbatim}
\usepackage{mathrsfs}
\usepackage{color}

\hypersetup{colorlinks,
linkcolor=blue,          citecolor=blue,        filecolor=blue,      urlcolor=blue           }

\def\be{\begin{equation}}
\def\ee{\end{equation}}
\def\bea{\begin{eqnarray}}
\def\eea{\end{eqnarray}}
\def\bse{\begin{subequations}}
\def\ese{\end{subequations}}

\def\be{\begin{eqnarray}}
\def\ee{\end{eqnarray}}

\begin{document}

\title{Scheme to Equilibrate the Quantized Hall Response of Topological Systems from Coherent Dynamics}
\author{Yong Xu$^1$}
\email{yongxuphy@tsinghua.edu.cn}
\author{Ying Hu$^{2,3}$}
\affiliation{$^1$ Center for Quantum Information, IIIS, Tsinghua University, Beijing 100084, PR China}
\affiliation{$^2$
State Key Laboratory of Quantum Optics and Quantum Optics Devices, Institute of Laser Spectroscopy, Shanxi University, Taiyuan, Shanxi 030006, China}
\affiliation{$^3$Collaborative Innovation Center of Extreme Optics, Shanxi University, Taiyuan, Shanxi 030006, China}

\begin{abstract}
Two-dimensional topologically distinct insulators are separated by topological gapless points, which exist
as Weyl points in three-dimensional momentum space.
Slowly varying parameters in the two-dimensional Hamiltonian across two distinct phases therefore necessarily
experiences the gap closing process, which prevents the intrinsic physical observable, the Hall response, from equilibrating.
To equilibrate the Hall response, engineered laser noises were introduced at the price of destroying the quantum coherence.
Here we demonstrate a new scheme to equilibrate the quantized Hall response from pure coherent dynamics as the Hamiltonian is slowly tuned from the topologically trivial to nontrivial regimes. We show the elements that affect the process of equilibration including the sequence when the electric field is switched on, its strength and the band dispersion of the final Hamiltonian. We further apply our method to Weyl semimetals in three dimensions and find the equilibrated Hall response  despite the underlying gapless band structure.
Our finding not only lays the theoretical foundation for observing the two-dimensional topological phase transition but also
for observing and controlling Weyl semimetals in ultracold atomic gases.
\end{abstract}
\maketitle

\begin{figure}[t]
\includegraphics[width=3.3in]{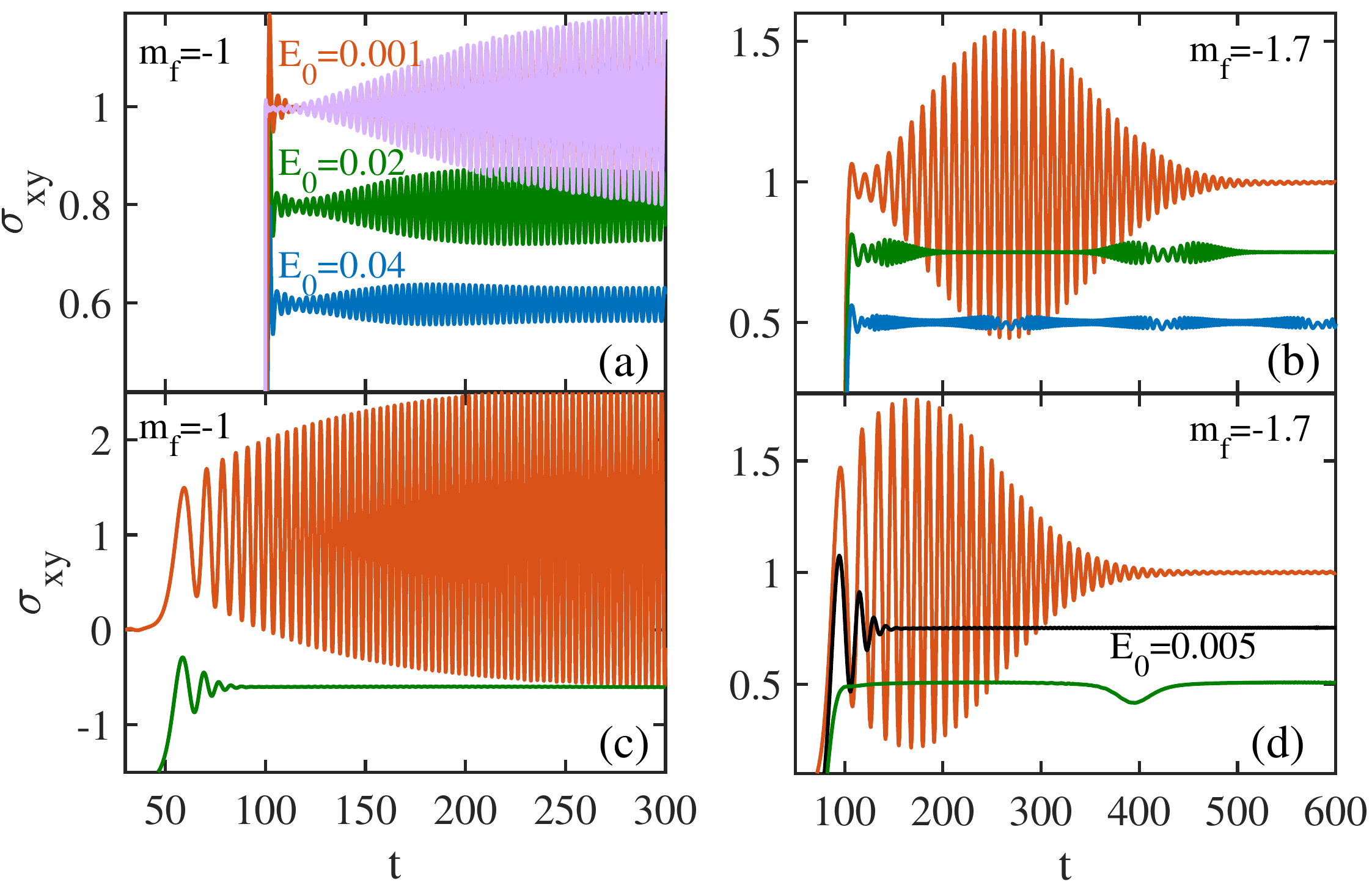}
\caption{(Color online) Coherent dynamics of the Hall response for various strength of electric field in two different protocols (see main text). Panels in the upper (lower) row correspond to the first (second) protocol.
Here $m_i=-2.5$, $v=0.02$ and $\tau_e=10$.
For visuality, we have shifted the Hall response downward for different $E_0$ from an asymptotically quantized value.
All curves are computed by the
exact numerical method except for the pink curve in panel (a), which is calculated based on the
perturbation approach.}
\label{Fig1}
\end{figure}

Recent realizations of the topological Haldane model~\cite{Esslinger2014,Sengstock2016} and the Chern insulator in the hyperfine spin space~\cite{Shuai2015,Shuai2017} with ultracold atomic gases provide an ideal platform to study the topological phase transition from non-equilibrium dynamics induced by either slow or rapid variation of
Hamiltonian parameters~\cite{Rigol2015,Bhaseen2015,Heyl2016,Kehrein2016,Wilson2016,Cooper2016b,Oktel2016,HuiZhai2017,
Rigol2017,ShuChen2017,Sengstock2017,Heyl2017PRB,Werner2017,Budich2017,Sengstock2018,ShuaiChen2018,Xiongjun2018,Jinglong2018,Ueda2018,WeiYi2018}.
In this context, one important and fundamental question is whether
slowly ramping the Hamiltonian through the topological phase transition, where the energy gap necessarily closes, leads to an equilibrated Hall response that signals the phase transition. Previous research shows that the Hall response indeed changes from non-equilibrium coherent dynamics but cannot reach equilibrium showing strong and long-lasting oscillations even for a slow parameter ramp, due to the inevitable non-adiabatic transitions through the gap closing point~\cite{YingHu2016,Goldman2017,Balseiro2018}. To equilibrate the Hall response,
engineered laser noises inducing pure dephasing in cold atom setups were introduced at the price of destroying quantum coherence in Ref.~\cite{YingHu2016}.
Instead, we will show a new scheme to equilibrate an asymptotically quantized Hall response of a Chern insulator
from pure \textit{coherent dynamics} as shown in Fig.~\ref{Fig1}.

The above issue also arises in the context of three-dimensional (3D) Weyl semimetals~\cite{VishwanathRMP}, which have seen a rapid development in a variety of  fields~\cite{volovik,Wan2011prb,Yuanming2011PRB,Burkov2011PRL,ZhongFang2011prl,Gong2011prl,LingLu2013NP,Yong2014PRL,Shengyuan2014PRL,
Tena2015RPL,Bo2015,Weng2015PRX,Xu2015,Lv2015,Lu2015,Xiaopeng2015,Yong2015PRL,Ueda2016,KTLaw2016,Yong2016typeii,ZhongWang2016PRL,Hubener2017,Lang2017}. Weyl semimetals feature \textit{gapless} band
structures characterized by the existence of pairs of Weyl points, which can be regarded as the topological
phase transition points between the topologically trivial and nontrivial insulators in momentum space. As a consequence, Weyl semimetals in their ground state are known to exhibit topological Hall response~\cite{Yuanming2011PRB,Burkov2011PRL}, which is not quantized, but rather determined by the location and number of Weyl points. In cold atom setups, creation and manipulation of Weyl points
can be readily achieved through the time-modulation of Hamiltonian parameters~\cite{Tena2015RPL,KTLaw2016,Yong2016typeii}. However,
whether the dynamical tuning of Hamiltonian parameters can give rise to the corresponding equilibrated Hall response has never been explored and is highly desired, which is nontrivial because of the gapless character of Weyl semimetals.

In this work, we demonstrate that an \textit{equilibrated} Hall response manifesting the topological property of the \textit{instantaneous} Hamiltonian can build up in \textit{non-equilibrium coherent dynamics}, despite topologically trivial time-evolved state.
We show this by slowly tuning a parameter of the experimentally realized Chern band from the topologically trivial to nontrivial regimes, thus driving the Hamiltonian through a phase transition at the gap closing point.
We find that the realization of an equilibrated topological Hall response relies on three elements: (i) whether the electric field is turned on before or after the gap closing point, (ii) the strength of electric field as compared to the ramp velocity, and (iii) the dispersion of the band structure of the final Hamiltonian.
We further show this equilibrated Hall response for a Weyl semimetal as pairs of Weyl points are created and their separation is continuously tuned by means of slow modifications of Hamiltonian parameters. Our findings thus provide the theoretical foundation for experimentally observing and controlling Weyl points in ultracold atomic gases.

\emph{Equilibrated Hall response of a Chern insulator}--- We start by considering a Chern insulator in the $x-y$ plane, motivated by its recent realization with ultracold atoms using Raman laser beams~\cite{Shuai2017}. The relevant time dependent Hamiltonian is $H[m_z(t)]=\sum_{\bf k}c_{\bf k}^\dag H[{\bf k}, m_z(t)]c_{\bf k}$ where
\begin{equation}
H[{\bf k}, m_z(t)]={\bm d}_{\bf k}(t)\cdot{\bm \sigma}
\label{Ham1}
\end{equation}
with $d_x=\sin(k_x)$, $d_y=\sin(k_y)$, $d_z=m_z(t)+\cos(k_x)+\cos(k_y)$ at lattice momentum ${\bf k}=(k_x,k_y)$, and ${\bf}{\sigma}=(\sigma_x,\sigma_y,\sigma_z)$ are the Pauli matrices. Here the energy is measured in units of hopping amplitude. In the cold-atom experiment, the mass parameter $m_z$ arises from the two-photon detuning and can be controlled. For static $m_z$, the ground state of Hamiltonian is in the topological nontrivial phase with Chern number $C=-\text{sgn}(m_z)$ for $-2<m_z<2$, otherwise, it is in the topological trivial phase with $C=0$.

We will be interested in the coherent dynamics of Hall response in an experimental relevant scenario, where the system is initially prepared in the topologically trivial ground state of Hamiltonian $H(m_i)$ with $m_i<-2$. We tune $m_z$ slowly from $m_i$ to $-2<m_f<0$ according to $m_z(t)=m_i+(m_f-m_i)(1-e^{-vt})$ with velocity $v$. This way, Hamiltonian $H[m_z(t)]$ changes its topology from trivial to topologically nontrivial, undergoing a transition at an energy gap closing point at critical time $t_c$. To probe Hall response, at time $t_e$, we ramp on an electric field in the $x$ direction as $E_x(t)=E_0(1-e^{-t/\tau_e})$~\cite{Bloch2015}, which can be generated via a homogeneous time-dependent synthetic gauge field, i.e., ${E_x}(t)=\partial_t {A}(t)$. The Hall current in the $y$ direction is computed by
\begin{equation}
J_{y}(t)=\frac{1}{2\pi}\int_{\textrm{BZ}}d{\bf k}\langle \psi_{\bf k}(t)|\partial_{k_y}H[{\bf k}^\prime(t),m_z(t)]|\psi_{\bf k}(t)\rangle,
\label{EqJxy1}
\end{equation}
where $|\psi_{\bf k}(t)\rangle$ denotes the instantaneous wavefunction at momentum ${\bf k}$, and the integration is over the first Brillouin zone (BZ). Note that the presence of electric field induces a shift in momentum via ${\bf k}^\prime(t)=[k_x+A(t),k_y]$. The Hall response is thus given by $\sigma_{xy}(t)=J_{y}(t)/E_x(t)$, measured in unit of $e^2/h$.

We will analyze and compare the coherent dynamics of Hall response in two different protocols for controlling the electric field: (1) We first vary the mass parameter without electric field $E_x(t)$. Then, some time after the transition of system through the critical point at time $t_c$, the $E_x(t)$ is turned on, i.e., $t_e\gg t_c$. (2) The $E_x(t)$ is turned on before the modulations of $m_z$, i.e., $t_e\ll t_c$. The numerical results for the coherent dynamics of Hall response in both protocols are shown in Fig.~\ref{Fig1}.

We find that, while a Hall response dynamically builds up after the Hamiltonian is ramped into a topological nontrivial regime, its equilibration to a topologically quantized value under coherent evolution crucially depends on two elements for both protocols: (i) dispersion of the energy band of the final Hamiltonian $H(m_f)$ and (ii) the magnitude of $E_0$. In more details, for $m_f=-1$ [Fig.~\ref{Fig1} (a) and (c)], where the corresponding energy band is flat along $k_x$ for $k_y=0$,
we see that the Hall response cannot reach equilibrium but rather exhibits strong and persistent oscillations for $E_0\ll v$ in both protocols [see red curves for $E_0=0.001$], as also found earlier~\cite{YingHu2016,Goldman2017,Balseiro2018}. Remarkably, when $E_0$ is increased to be comparable to the ramp velocity [e.g., see green curves for $E_0=0.02$], such irregularities disappear after a few oscillations in protocol (2), as opposed to its counterpart in
protocol (1) where the oscillations are only moderately suppressed. When the underlying energy band becomes increasingly dispersive, such as for $m_f=-1.7$ [Figs.~\ref{Fig1} (b), (d)], we see that the oscillations of the Hall response generically damps out at long times even for weak $E_0$ in both protocols, and increasing $E_0$ can significantly reduce the time for equilibration. These findings show that an equilibrated asymptotically quantized Hall response can build up from coherent dynamics, despite the non-adiabatic passage through the gap closing.

To gain insight into above dynamical behavior of Hall response, we first analyze protocol (1) using the perturbation approach. In this case, the energy gap closes at momentum ${\bf k}_c=(0,0)$ at time $t_c$. Then at time $t_e$ when $E_x(t)$ is turned on, for simplicity and to capture the essential physics, we assume that $H(m_f)$ has been reached and the state can be described by
$|\psi_{\bf k}(t_e)\rangle=\sqrt{1-p({\bf k})}e^{i D_1({\bf k})}|u_-({\bf k})\rangle+e^{i\theta({\bf k})}e^{-i D_1({\bf k})}\sqrt{p({\bf k}) }|u_+({\bf k})\rangle$ for ${\bf k}\neq 0$,
where
$|u_{+(-)}({\bf k}) \rangle$ denotes the excited (ground) state of $H({\bf k},m_f)$, $\theta({\bf k})$ is the relative phase, and
$D_1({\bf k})=\int_{0}^{t_e} \epsilon_{+}({\bf k},m_z(t^\prime))dt^\prime$ is the dynamical phase with $\epsilon_{+(-)}({\bf k},m_z(t))$ the
eigenenergy of Hamiltonian $H({\bf k},m_z(t))$. Moreover, $p({\bf k})$ denotes the number of excitations created during the non-adiabatic passage through the gap closing. At times $t>t_e$, where the gap has been reopened, using time-dependent perturbation theory~\cite{XiaoRMP}, the evolution of $|u_\lambda({\bf k})\rangle$ can be approximated by ($\hbar\equiv 1$)
$
|\Phi_{\lambda}({\bf k}^\prime,t)\rangle = e^{-i\int_{t_e}^t \epsilon_\lambda({\bf k}^\prime)dt^\prime}e^{i\gamma_{\lambda}(t)}
\times \Big[|u_\lambda({\bf k}^\prime)\rangle-iE_0\frac{\langle u_{\bar\lambda}({\bf k}')|\partial_{k_x} |u_{\lambda}({\bf k}')\rangle}{2\lambda\epsilon_{+}({\bf k}')}|u_{\bar\lambda}({\bf k}^\prime)\rangle\Big] %\label{eq:u}
$.
Here $\bar\lambda\neq \lambda$ and $\gamma_{\lambda}(t)$ is associated with the Berry phase in the $\lambda$ band. Therefore, after ramping on the electric field, the state evolves from $|\psi_{\bf k}(t_e)\rangle$ into:
$
|\psi_{\bf k}(t>t_e)\rangle=\sqrt{1-p({\bf k})}e^{i D_1({\bf k})}|\Phi_-({\bf k}^\prime,t)\rangle+e^{i\theta({\bf k})}
e^{-i D_1({\bf k})} \sqrt{p({\bf k})}|\Phi_+({\bf k}^\prime,t)\rangle.
$
Substitution of this ansatz into Eq.~(\ref{EqJxy1}) then gives the Hall current,
which well captures the dynamical behavior of Hall response at times $t> t_e$, as evidenced by the pink curve in Fig.~\ref{Fig1}(a).

The Hall current contains three contributions, i.e.,
\begin{equation}
J_{y}(t)=J_{\textrm{Hall}}(t)+J_{\textrm{Dis}}(t)+J_{\textrm{Osc}}(t).
\end{equation}
Here, $J_{\textrm{Hall}}(t)=(E_0/2\pi)\int_{\textrm BZ}d{\bf k}[p({\bf k})\Omega_{+}({\bf k}^\prime)+(1-p({\bf k}) \Omega_{-}({\bf k}^\prime)]$ describes the weighted anomalous Hall current which has topological origin, with $\Omega_{\pm}$ labeling the instantaneous Berry curvature for the upper (lower) band of $H({\bf k}', m_f)$. In the limit of vanishing ramp velocity, $J_{\textrm{Hall}}(t)/E_0$ approaches the Chern number of $H(m_f)$. The current
$J_{\textrm{Dis}}=(1/2\pi)\int_{\textrm BZ}d{\bf k}[p({\bf k})\partial_{k_y}\epsilon_{+}({\bf k}^\prime)+(1-p({
 \bf k}))\partial_{k_y}\epsilon_{-}({\bf k}^\prime)]$ arises from the weighted band velocity, which exactly vanishes because of the underlying symmetry in our system, i.e., $\epsilon_\lambda(k_x,k_y)=\epsilon_\lambda(k_x,-k_y)$. The current $J_{\textrm{Osc}}(t)$ comes from the coherent superpositions between the upper and lower bands, i.e.,
\begin{equation}
J_{\textrm{Osc}}(t)=\textrm{Re}\int_{\textrm{BZ}} d{\bf k}\sqrt{p_{\bf k}(1-p_{\bf k} )} e^{i[\theta({\bf k})-D_2({\bf k}^\prime)]}\Gamma(t),
\label{eq:J3}
\end{equation}
where $D_2({\bf k}^\prime)=2\int_{t_e}^t dt' \epsilon_+({\bf k}^\prime)$ and $\Gamma(t)=\frac{1}{\pi}e^{-2iD_1({\bf k})}e^{i(\gamma_{+}-\gamma_{-})} \left[g_{12}-2iE_0 h_{12}\frac{\partial_{k_y}\epsilon_+({\bf k}^\prime,m_f) }{2\epsilon_+({\bf k}^\prime,m_f)}\right]$
with $g_{12}({\bf k}^\prime)=\langle u_{-}({\bf k}^\prime)|\partial_{k_y}H({\bf k}^\prime,m_f)|u_{+}({\bf k}^\prime)\rangle$, and $h_{12}=\langle\partial_{k_x}u_{-}({\bf k}^\prime)|u_{+}({\bf k}^\prime)\rangle$. Obviously, $J_{\textrm{Osc}}(t)$ is responsible for the oscillation as we detail below.

\begin{figure}[t]
\includegraphics[width=3.3in]{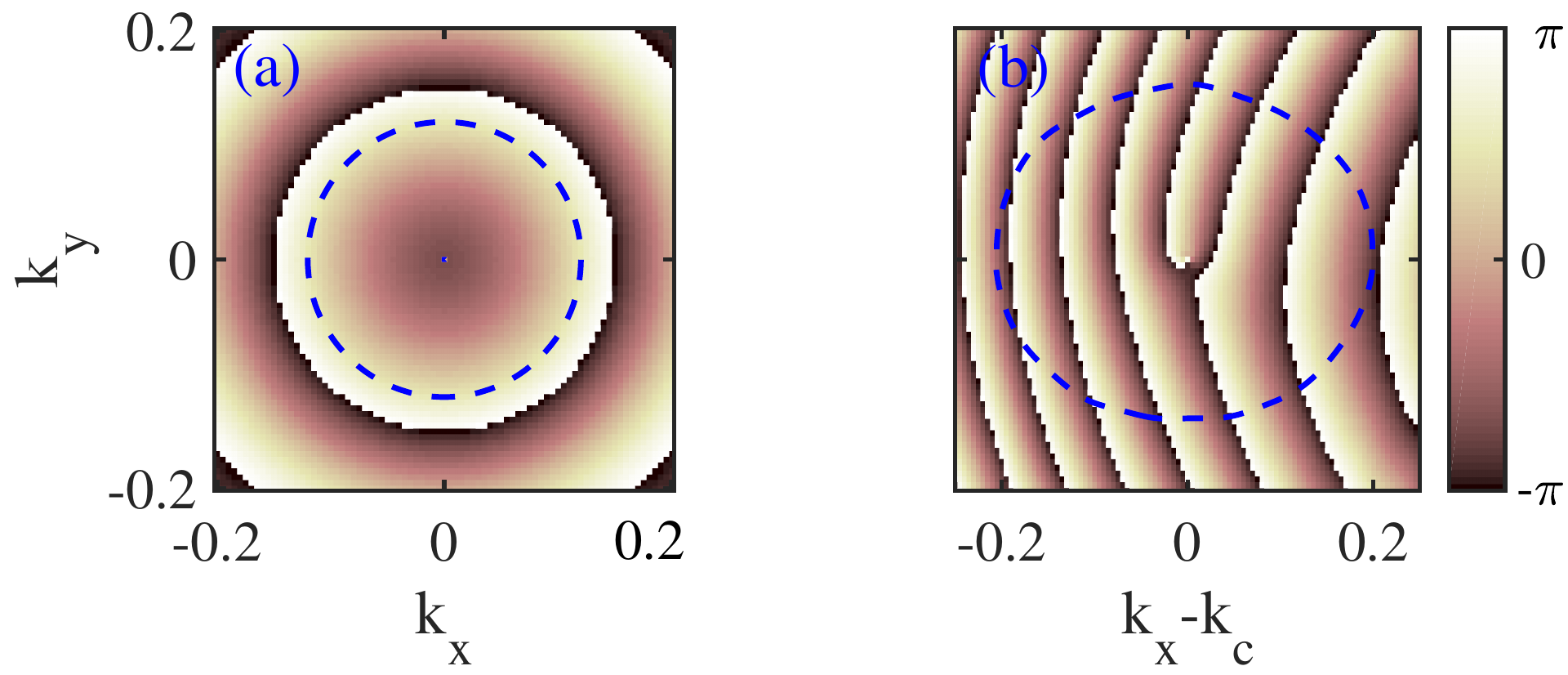}
\caption{(Color online) Profiles of $\theta({\bf k})$ near the gap closing point ${\bf k}_c$ (for ${\bf k}\neq{\bf k}_c$)
in the $k_x-k_y$ plane for (a) protocol (1) with $k_c=0$ and (b) protocol (2) with $k_c=-A(t_c)$ and $E_0=0.02$.
At lattice momenta depicted by the dashed blue line, the corresponding population in the excited state is $0.1$.}
\label{Fig2}
\end{figure}

Equation (\ref{eq:J3}) allows us to understand the effect of electric field on the oscillations and their damping in the coherent dynamics of Hall response as shown in Figs.~\ref{Fig1}(a) and (b).
In our analysis we will focus on the term involving $g_{12}$ while ignoring that involves $h_{12}$, which is small due to gapped energy as is numerically verified. For a slow ramp, the excitations occur in a very narrow region near the gap-closing point at ${\bf k}_c=(0,0)$, which can be described in the context of Landau-Zener (LZ) physics that gives $p({\bf k})\approx e^{-\pi k^2/v_{\textrm{LZ}}}$, with $v_{\textrm{LZ}}=(m_f+2)v$. Hence the dominant contribution to $J_{\textrm{Osc}}(t)$ comes from the momenta $|{\bf k}|<\sqrt{v_{\textrm{LZ}}/\pi}$. In this regime, in the limit of weak electric field $E_0\rightarrow 0$, one can expand $g_{12}$ in terms of $A$ on the time scale $t\ll (\sqrt{v_{\textrm{LZ}}/\pi})/E_0$, and the main contribution to $J_{\textrm{Osc}}(t)$ comes from the first term, i.e., $g_{12}\propto A=E_0(t-t_e)$. Therefore, for sufficiently weak electric field such as in $E_0=0.001$, the amplitude of oscillating Hall response will always undergo an initial increase with time, as shown in Figs.~\ref{Fig1}(a) and (b).

In the long time limit when $\sqrt{v_{\textrm{LZ}}/\pi} \ll A(t)<2\pi/E_0$, which is quickly fulfilled as in the case when $E_0=0.02$, we find that the oscillations persist for $m_f=-1$ while
can be damped out for $m_f\neq -1$ as shown in Figs.~\ref{Fig1}(a) and (b). The damping is caused by the significant difference for the energy spectrum $\epsilon_+({\bf k}')$ along the $k_x$ direction, which is strongly modified by the presence of large $A(t)$.
To gain some intuitive understanding, let us keep only the dominant term responsible for the strong damping, $D_2({\bf k}^\prime)$, and approximate $J_{\textrm{Osc}}(t)$ by
$J_{\textrm{Osc}}(t)\sim 2\textrm{Re}\int d{\bf k}\sqrt{p({\bf k})(1-p({\bf k}) )} e^{2i\epsilon_{-}({\bf k}+{\bf k}_e,m_f)t^\prime}$ with $t^\prime =t-t_e$,
where we have taken the leading term $g_{12}\approx 1$ and ignored the time independent phase contribution and the Berry phase, which is irrelevant due to its small variation in momentum space. To see the effect of spectrum difference, let us fix ${\bf k}_e$ (which is driven by the electric field by $k_{ex}=E_0 t^\prime$), e.g., $k_{ex}=\pi/2$, where dispersion exhibits the largest derivative along $k_x$ for $k_y=0$ and $|m_f| \neq 1$. We find that the oscillations decays exponentially
for $m_f=-1.5$ and $v=0.02$. However, when $m_f=-1$, we find $J_{Osc} \sim \sqrt{c t^\prime}/(1-ict^\prime)$ with $c=2v_{LZ}/\pi$ in the long time limit, which decays very slowly,
almost leading to a persistent oscillation, consistent with Fig.~\ref{Fig1}(a), because of flat energy dispersion for $k_y=0$.
Interestingly, we further see that the oscillations of Hall currents revive as it approaches a period of $2\pi/E_0$ because Bloch oscillation occurs [see blue curve in Fig.~\ref{Fig1}(b)].
This also implies that the oscillations of Hall currents cannot be completely damped out if the relevant time scale for damping is $>\pi/E_0$.

Above analysis guides our intuition into the remarkably equilibrated Hall response in protocol (2). As $E_x(t)$ is present initially before $m_z(t)$ is varied, the energy gap closing point is shifted to ${\bf k}_c=[-A(t_c),0]$. This motivates us to consider the form of Eq.~(\ref{eq:J3}) with the replacement $p({\bf k})\rightarrow p({{\bf k}+{\bf k}_c}) $ and $\theta({\bf k})\rightarrow \theta({\bf k}+{\bf k}_c)$. When $E_0=0.02$, we find that $\theta({\bf k}+{\bf k}_c)$ exhibits rapid variations near ${\bf k}_c$ along the $x$ direction [see Fig.~\ref{Fig2}(b) calculated by the exact numerical method], in sharp contrast to the counterpart of protocol (1) [see Fig.~\ref{Fig2}(a)], where $\theta({\bf k})$ varies slowly.
The rapid variations introduce the strong damping when the integration over momentum space is performed, leading to a rapid equilibration of Hall response in protocol (2) reflecting the topology of the instantaneous Hamiltonian, even for $m_f=-1$.

\emph{Equilibrated Hall response of Weyl semimetals}--- We now extend our analysis to three dimensional (3D) Weyl semimetals, which can exhibit anomalous Hall effects. We consider a system of atoms in a 3D lattice described by the Hamiltonian (see SM for realization scheme)
\begin{equation}\label{eq:Hc}
H_{C}=\frac{{\bf p}^{2}}{2m}-\sum _{\nu=x,y,z}V_{\nu}\cos^{2}(k_{L\nu}r_{\nu})+m_{z}\sigma_{z}+V_{\textrm{SO}}.
\end{equation}
Here $m$ is the mass of atoms, ${\bf p}$ is the momentum operator, $V_{\textrm{SO}}=M_{y}\sigma_x-M_{x}\sigma_y$ describes the nondiagonal optical lattices with $M_{x}=\Omega_{SO}\sin(k_{Lx}r_{x})\cos(k_{Ly}r_{y})\cos(k_{Lz}r_{z})$ and
$M_{y}=\Omega_{\textrm{SO}}\sin(k_{Ly}r_{y})\cos(k_{Lx}r_{x})\cos(k_{Lz}r_{z})$. Further, the diagonal lattice potential in direction $\nu$ is characterized by the strength $V_\nu>0$ and the period $a_\nu=\pi/k_{L\nu}$. Hamiltonian (\ref{eq:Hc}) can be recast into the following tight-binding form (see SM for derivation), i.e.,
$H_{\textrm{TB}}=\sum_{{\bf x}}[-\sum_{\nu}(J_{\nu}\hat{c}_{{\bf x}}^{\dagger}\hat{c}_{{\bf x}+a_{\nu}{\bf e}_{\nu}}+\textrm{H.c}.)+m_{z}\hat{c}_{{\bf x}}^{\dagger}\sigma_{z}\hat{c}_{{\bf x}}+(-1)^{j_{x}+j_{y}+j_{z}}J_{\textrm{SO}}(
\hat{c}_{{\bf x}}^{\dagger}\sigma_{y}\hat{c}_{{\bf x}+a_{x}{\bf e}_{x}}-\hat{c}_{{\bf x}}^{\dagger}\sigma_{x}\hat{c}_{{\bf x}+a_{y}{\bf e}_{y}}+\textrm{H.c}.)],
$. Here $\hat{c}_{{\bf x}}^{\dagger}=\left(\begin{array}{cc}
\hat{c}_{{\bf x}\uparrow}^{\dagger} & \hat{c}_{{\bf x}\downarrow}^{\dagger}\end{array}\right)$ with $\hat{c}_{{\bf x}\sigma}$ ( $\hat{c}_{{\bf x}\sigma}^{\dagger}$) annihilating (creating) an atom
with spin $\sigma$ at ${\bf x}=\sum_{\nu}j_{\nu}a_{\nu}{\bf e}_{\nu}$. The corresponding Bloch Hamiltonian at lattice momentum ${\bf k}=(k_x,k_y,k_z)$ can be written as
\begin{equation}\label{eq:H3D}
H_{\textrm{3D}}({\bf k})=m_{z}\sigma_{z}-h_{t}\tau_{x}+\tau_{y}(d_{y}\sigma_{x}-d_{x}\sigma_{y}),
\end{equation}
where $\tau_\nu$ denotes the Pauli matrix describing the sublattices,
$h_{t}=-2\sum_{\nu=x,y,z}J_{\nu}\cos(k_{\nu}a_{\nu})$, $d_{x}=2J_{\textrm{SO}}\sin(k_{x}a_{x})$
and $d_{y}=2J_{\textrm{SO}}\sin(k_{y}a_{y})$. The eigenenergy corresponding to Hamiltonian (\ref{eq:H3D}) is
$E_\pm(\lambda)=\pm\sqrt{d_{x}^{2}+d_{y}^{2}+(h_{t}-\lambda m_{z})^{2}}
$ with $\lambda=\pm 1$. Below we will assume  $J_\nu=J$ for convenience.

\begin{figure}[t]
\includegraphics[width=3.2in]{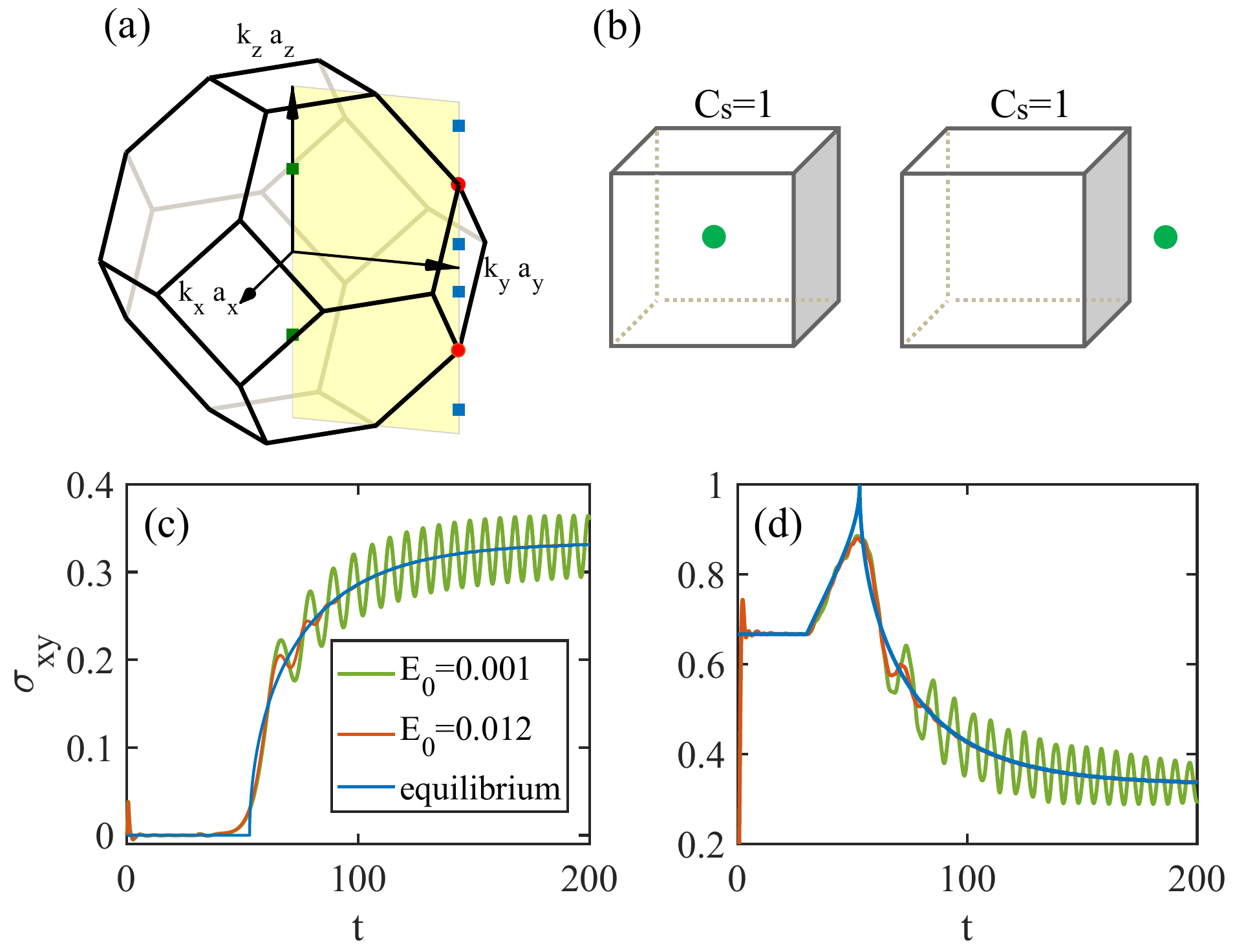}
\caption{(Color online) (a) The first Brillouin zone with a pair of Dirac points,
one and two pairs of Weyl points, denoted by the solid red circles, green and blue squares, respectively.
(b) Schematics illustrating that the Chern number of states remains unchanged on a closed surface when
a Weyl point moves out of the surface. Evolution of the Hall response over time as $m_z$ varies
slowly (c) from $-7J$ to $-5J$ and (d) from $-3J$ to $-J$. Here the unit of time is $\hbar/J$.}
\label{Fig3}
\end{figure}

The semimetal (\ref{eq:H3D}) exhibits a very rich phase diagram. When $|m_z|>6J$, the Weyl semimetal is in the trivial insulating phase. When $2J<|m_z|<6J$, the system is in the topological phase featuring one pair of Weyl points, whereas for $|m_z|<2J$ and $m_z\neq 0$, the semimetal has two pairs of Weyl points. For $m_z=0$, the system is
the Dirac semimetal with two Dirac points~\cite{footnote1}. The locations of Weyl points and Dirac points
in momentum space are displayed in Fig.~\ref{Fig3}(a). If we choose a closed surface
enclosing a Weyl point as shown in Fig.~\ref{Fig3}(b), we find that the Chern number of
states on the surface remains unchanged over time even if the Weyl point moves out of the
surface, implying the Chern number of states does not reflect the Chern number of the Hamiltonian.

In Figs.~\ref{Fig3}(c) and (d), we show the coherent dynamics of Hall response in the $y$ direction for the considered Weyl semimetal~\cite{footnote2}, when an electric field $E_x(t)$ is switched on before $m_z(t)$ is slowly tuned. Importantly, we see that the Hall response at $E_0=0.012$ is finally equilibrated to the value of its equilibrium counterpart dictated by the instantaneous Hamiltonian as we slowly tune the system from a topological trivial insulator to a Weyl semimetal phase with two Weyl points (see Fig.~\ref{Fig3}(c)) and from a phase with two Weyl points to another phase with four points (see Fig.~\ref{Fig3}(d)); this is in sharp contrast to the case for $E_x=0.001$ where the Hall response cannot be equilibrated but rather exhibits strong oscillations, as we have discussed earlier.

In summary, we have demonstrated a new scheme to equilibrate the quantized Hall response from pure coherent dynamics
when a 2D insulator is ramped from a topologically trivial into nontrivial regime. We further apply our strategy to a 3D Weyl semimetal and find the equilibrated Hall response as the number and location of Weyl points are slowly tuned.
Our findings not only pave the way for observing the 2D topological phase transition but also for observing and controlling Weyl semimetals in ultracold atomic gases.

\begin{acknowledgments}
\textit{Acknowledgement:} We thank B. Wu, D.-L. Deng, Y.-K. Wu, Z.-X. Gong and T. Qin for helpful discussions.
Y. X. is supported by
Tsinghua start up program and National Thousand-Young-Talents Program. Y. H. acknowledges support from the National Thousand-Young-Talents Program, the National Key Research and Development Program of China (Grant Nos. 2016YFA0301700, 2017YFA0304203), and Changjiang Scholars and Innovative Research Team in University of Ministry of Education of China (Grant No. IRT13076).
\end{acknowledgments}

\begin{widetext}
\section{Supplemental Material}
%%%%%%%%%% Prefix a "S" to all equations, figures, tables and reset the counter %%%%%%%%%%
\setcounter{equation}{0} \setcounter{figure}{0} \setcounter{table}{0} %
\renewcommand{\theequation}{S\arabic{equation}} \renewcommand{\thefigure}{S%
\arabic{figure}} \renewcommand{\bibnumfmt}[1]{[S#1]} \renewcommand{%
\citenumfont}[1]{S#1}
%%%%%%%%%% Prefix a "S" to all equations, figures, tables and reset the counter %%%%%%%%%%

In the supplemental material, we will provide the details for calculation of the phases $\theta({\bf k}+{\bf k}_c)$ induced
by the electric field and
propose an experimental scheme for realization of a Weyl semimetal described by the continuous model in Eq. (5)
in the main text and derive its tight-binding Hamiltonian.

First, to calculate the phase $\theta({\bf k}+{\bf k}_c)$ induced by the electric field,
we expand the state in the basis of instantaneous eigenstates $|u_{\lambda}({\bf k}^\prime,m_z(t))\rangle$ as
\begin{equation}
|\psi_{\bf k}(t)\rangle=\sum_{\lambda=\pm}\alpha_\lambda({\bf k},t) A_\lambda ({\bf k},t)|u_\lambda({\bf k}^\prime,m_z(t))\rangle,
\label{psiInst}
\end{equation}
where $|u_{\lambda}({\bf k}^\prime,m_z(t))\rangle$ satisfies $H({\bf k}^\prime,m_z(t))|u_{\lambda}({\bf k}^\prime,m_z(t))\rangle=\epsilon_\lambda({\bf k}^\prime,m_z(t)) |u_{\lambda}({\bf k}^\prime,m_z(t))\rangle$ with $\epsilon_\lambda({\bf k}^\prime,m_z(t))=\lambda\sqrt{\sum_{\nu=x,y,z} d_\nu^2({\bf k}^\prime,m_z(t))}$ and $\lambda=\pm 1$, $A_\lambda ({\bf k},t)=e^{-i\int_0^t \epsilon_\lambda({\bf k}^\prime,m_z(t^\prime))dt^\prime}e^{i\gamma_{\lambda}(t)}$
corresponding to the dynamical and Berry phases $\gamma_{\lambda}(t)=i\int_0^t dt^\prime \langle u_\lambda({\bf k}^\prime,m_z(t^\prime))|\partial_{t^\prime} u_\lambda({\bf k}^\prime,m_z(t^\prime))\rangle$, respectively.
Plugging Eq.~(\ref{psiInst}) into the Schr\"odinger equation yields
\begin{eqnarray}
&&\partial_t\alpha_-=-f(t) \alpha_+ A_+/A_-,
\label{Eqalphas1}
\\
&&\partial_t\alpha_+=f(t)^* \alpha_- A_-/A_+,
\label{Eqalphas2}
\end{eqnarray}
where $f(t)=\langle u_-({\bf k}^\prime,m_z(t))|\partial_t u_+({\bf k}^\prime,m_z(t))\rangle$. Supposing that all atoms are initialized to
the lower band, we can calculate the time evolution of $\alpha_\lambda({\bf k},t)$, obtaining $\theta({\bf k})=\textrm{angle}(\alpha_+({\bf k},t>t_c)/\alpha_-({\bf k},t>t_c))$.

Second, to implement the continuous model in Eq. (5) in the main text, we only need to slightly modify
our previous scheme for realization of a dynamical 4D Weyl nodal ring~\cite{Yong2017Arxiv}.
We refer the reader to Figs. (3)(c-d) for a
laser configuration setup, where two sets of Raman laser beams are utilized to generate the
off-diagonal spin-dependent optical lattices. Each set includes two pairs of Raman laser beams.
In the first set, for one pair, the Rabi frequencies are: [$\bar{\Omega} _{1}=-\bar{\Omega}_{10}\cos
(k_{Ly}r_{y})e^{-ik_{Lz}r_{z}/2}$, $\bar{\Omega} _{2}=i\bar{\Omega}_{20}\sin
(k_{Lx}r_{x})e^{ik_{Lz}r_{z}/2}$], and for the other pair, they are
[$\bar{\Omega} _{1}^{\prime }=-\bar{\Omega} _{10}\cos
(k_{Ly}r_{y})e^{ik_{Lz}r_{z}/2}$, $\bar{\Omega} _{2}^{\prime }=i\bar{\Omega}
_{20}\sin (k_{Lx}r_{x})e^{-ik_{Lz}r_{z}/2}$].
In the second set, for one pair, the Rabi frequencies are:
[$\tilde{\Omega} _{1}=\tilde{\Omega} _{10}\sin
(k_{Ly}r_{y})e^{-ik_{Lz}r_{z}/2}$, $\tilde{\Omega} _{2}=\bar{\Omega}
_{10}\cos (k_{Lx}r_{x})e^{ik_{Lz}r_{z}/2}$], and for the other pair, they are
[$\tilde{\Omega} _{1}^{\prime }=\tilde{\Omega} _{10}\sin
(k_{Ly}r_{y})e^{ik_{Lz}r_{z}/2}$, $\tilde{\Omega} _{2}^{\prime }=\bar{\Omega}
_{10}\cos (k_{Lx}r_{x})e^{-ik_{Lz}r_{z}/2}$]. We also require another laser beam to
create an optical lattice along $z$. Using this scheme, we can achieve the Hamiltonian
in Eq. (7).

To obtain the continuous model's tight-binding Hamiltonian, let us write down the many-body Hamiltonian
using the field operator
\begin{equation}
H_{II}=\int d\mathbf{r}\hat{\psi}^{\dagger }(\mathbf{r})H_{C}\hat{\psi}(\mathbf{r%
}),
\label{HSec}
\end{equation}%
where $%
\hat{\psi}(\mathbf{r})=[%
\begin{array}{cc}
\hat{\psi}_{\uparrow }(\mathbf{r}) & \hat{\psi}_{\downarrow }(\mathbf{r})%
\end{array}%
]^{T}$ with $\hat{\psi}_{\sigma }(\mathbf{r})$ [$\hat{\psi}_{\sigma
}^{\dagger }(\mathbf{r})$] being a field operator destroying (creating)
a particle located at ${\bf r}$ with spin $\sigma $ ($\sigma =\uparrow ,\downarrow $).
The anti-commutation or commutation relation $[\hat{\psi}_{\sigma }(\mathbf{r}),%
\hat{\psi}_{\sigma ^{\prime }}^{\dagger }(\mathbf{r}^{\prime })]_{\pm
}=\delta _{\sigma \sigma ^{\prime }}\delta (\mathbf{r}-\mathbf{r}^{\prime })$
are required to be respected for fermionic ($+$) or bosonic operators ($-$), respectively.

We approximately expand the field operator as
\begin{equation}
\hat{\psi}_{\sigma }(\mathbf{r})\approx\sum_{{\bf x},\sigma }W_{{\bf x}}({\bf r})\hat{c}%
_{{\bf x},\sigma },
\label{EWn}
\end{equation}%
where $\hat{c}_{{\bf x},\sigma}$ is the annihilation operator for a particle with spin $\sigma$
located at the site ${\bf x}$, which satisfies the anti-commutation or commutation relation
$[\hat{c}_{{\bf x},\sigma},\hat{c}_{{\bf x}^\prime,\sigma^\prime}^\dagger]_{\pm}=\delta_{{\bf x},{\bf x}^\prime}
\delta_{\sigma,\sigma^\prime}$
for fermionic ($+$) or bosonic ($-$) atoms, respectively, and
$W_{\bf x}({\bf r})$ is the Wannier function for the lowest band of the Hamiltonian with
$h_z=V_{SO}=0$, which is located at the site ${\bf x}=\sum_\nu j_\nu a_\nu {\bf e}_\nu$
with $\nu=x,y,z$.

With the aid of Eq.~(\ref{EWn}), we can obtain the following tight-binding Hamiltonian by keeping only
the nearest-neighbor hopping terms (see Ref.~\cite{SMYong2016PRA,SMYong2016typeii} for the detailed
derivation and verification for its validity),
\begin{eqnarray}
H_{TB}=&&\sum_{{\bf x}}\left[-\sum_{\nu}(J_{\nu}\hat{c}_{{\bf x}}^{\dagger}\hat{c}_{{\bf x}+a_{\nu}{\bf e}_{\nu}}+H.c.)+m_{z}\hat{c}_{{\bf x}}^{\dagger}\sigma_{z}\hat{c}_{{\bf x}}\right] \nonumber \\
&&+\sum_{{\bf x}}(-1)^{j_{x}+j_{y}+j_{z}}J_{SO}\left(\hat{c}_{{\bf x}}^{\dagger}\sigma_{y}\hat{c}_{{\bf x}+a_{x}{\bf e}_{x}}-\hat{c}_{{\bf x}}^{\dagger}\sigma_{x}\hat{c}_{{\bf x}+a_{y}{\bf e}_{y}}+H.c.\right),
\end{eqnarray}
where $\hat{c}^\dagger_{\bf x}=(\hat{c}^\dagger_{{\bf x},\uparrow},\hat{c}^\dagger_{{\bf x},\downarrow})$.
In the basis of
$\Psi({\bf k})^{\dagger}=(\begin{array}{cccc}
e^{-ik_{x}a_{x}}\hat{A}_{{\bf k}\uparrow}^{\dagger} & e^{-ik_{x}a_{x}}\hat{A}_{{\bf k}\downarrow}^{\dagger} & \hat{B}_{{\bf k}\uparrow}^{\dagger} & \hat{B}_{{\bf k}\downarrow}^{\dagger}\end{array})$ where $A$ and $B$
correspond to two sublattices, the Hamiltonian can be written as in the momentum space
$H_{TB}=\sum_{\bf k} \Psi({\bf k})^{\dagger} H_{3D}({\bf k})\Psi({\bf k})$, where $H_{3D}({\bf k})$ is the Hamiltonian (5)
in the main text.
Applying the transformation $\hat{a}_{{\bf x} \uparrow}=(-1)^{j_x+j_y+j_z}\hat{c}_{{\bf x}\uparrow}$ and
$\hat{a}_{{\bf x}\downarrow}=\hat{c}_{{\bf x}\downarrow}$ reduces the model to the form
\begin{eqnarray}
H_{TB}^\prime & =\sum_{{\bf x}}\left[\sum_{\nu}(J_{\nu}\hat{a}_{{\bf x}}^{\dagger}\sigma_{z}\hat{a}_{{\bf x}+a_{\nu}{\bf e}_{\nu}}+H.c.)+m_{z}\hat{a}_{{\bf x}}^{\dagger}\sigma_{z}\hat{a}_{{\bf x}}-J_{SO}\left(i\hat{a}_{{\bf x}}^{\dagger}\sigma_{y}\hat{a}_{{\bf x}+a_{y}{\bf e}_{y}}+i\hat{a}_{{\bf x}}^{\dagger}\sigma_{x}\hat{a}_{{\bf x}+a_{x}{\bf e}_{x}}+H.c.\right)\right],
\end{eqnarray}
where $\hat{a}^\dagger_{\bf x}=(\hat{a}^\dagger_{{\bf x},\uparrow},\hat{a}^\dagger_{{\bf x},\downarrow})$. The
lattice structure becomes simple orthorhombic from a rocksalt lattice structure. Using the Fourier transformation,
we can write this Hamiltonian in the momentum space as $H_{TB}^\prime=\sum_{\bf k}\hat{a}_{\bf k}^\dagger H_{3D}^\prime({\bf k})\hat{a}_{\bf k}$, where $\hat{a}^\dagger_{\bf k}=(\hat{a}^\dagger_{{\bf k},\uparrow},\hat{a}^\dagger_{{\bf k},\downarrow})$ and $H_{3D}^\prime({\bf k})$ is given in the footnote of the main text.

\end{widetext}

\end{document}